\documentstyle[nato,epsf,namedreferences]{crckapb}

\input psfig

\font\FermiPPTfont=cmssbx10 scaled 1440
\font\FermiSmallfont=cmssq8 scaled 1200

\def\FNALpptheadhighest#1#2{
\null \vskip -3.75truein
\centerline{\hbox to 7.5truein {
\vbox to 1in{\vfill 
             \hbox{\psfig{figure=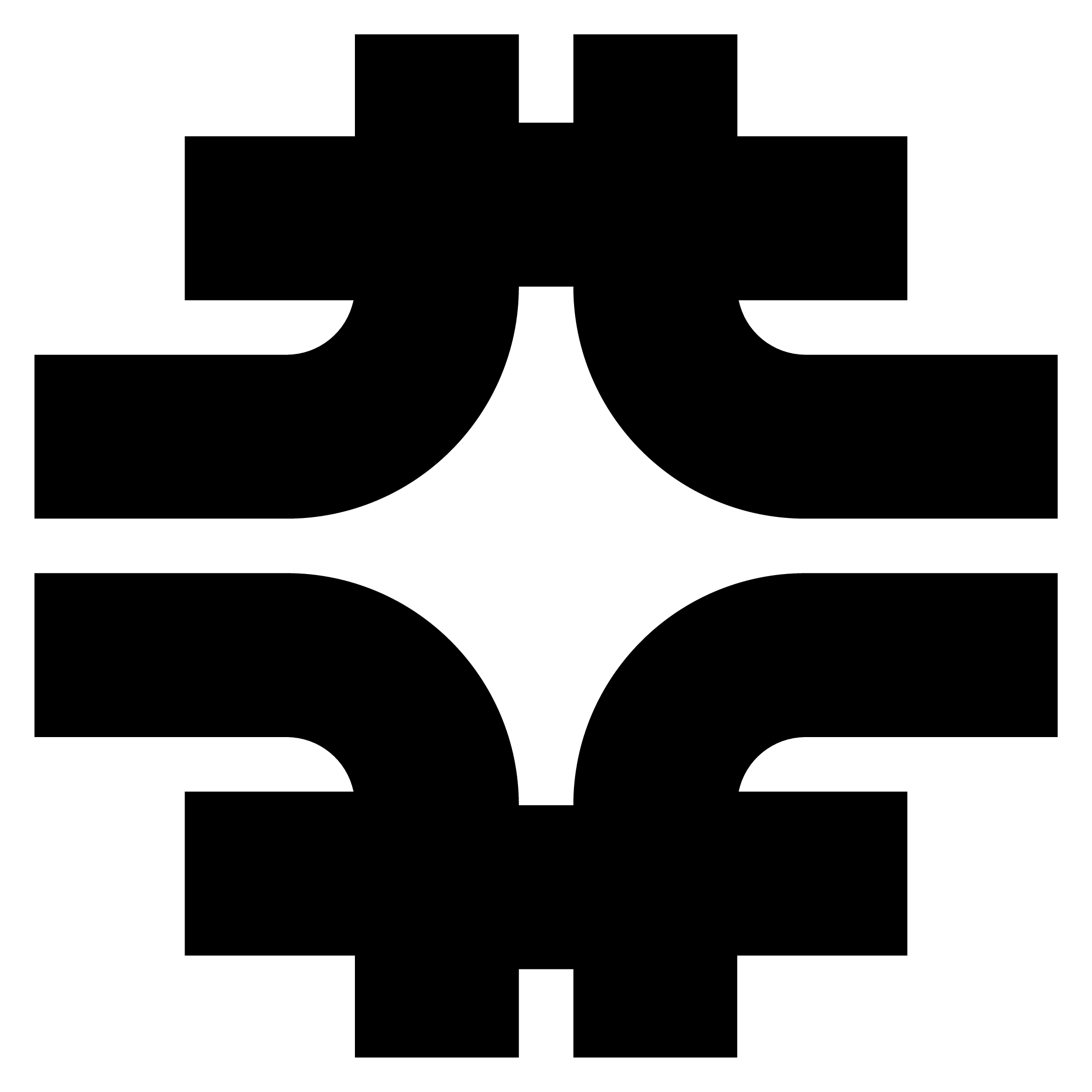,height=1.5cm,clip=}} 
             \vfill }
\hskip 1em
\vbox to 1in{\vfill
             \hbox{{\FermiPPTfont Fermi National Accelerator Laboratory}}
             \vfill}
\hfill
\vbox to 1in {\vfill \FermiSmallfont
              \hbox{#1}
              \hbox{#2}
              \vfill}
}}\vskip +2.75truein}%FNALpptheadhighest

\newcommand{\s}{\scriptscriptstyle}
\newcommand{\rmb}{{\rm b}}

\def\timedate{ {\tt
\count215=\time \divide\count215 by60  \number\count215
\multiply\count215 by-60 \advance \count215 by\time :\number\count215 \space
\number\day\space
\ifcase\month\or January\or February\or March\or April\or May\or June\or July
\or August\or September\or October\or November\or December\fi\space\number\year
}}
\def\s {\scriptscriptstyle}
\def\ccdot{{\hskip-0.7pt\cdot\hskip-0.7pt}}

\def\sqr#1#2{{\vcenter{\hrule height.#2pt
              \hbox{\vrule width.#2pt height#1pt \kern#1pt \vrule width.#2pt}
              \hrule height.#2pt}}}

\def\mathrelfun#1#2{\lower3.6pt\vbox{\baselineskip0pt\lineskip.9pt
  \ialign{$\mathsurround=0pt#1\hfil##\hfil$\crcr#2\crcr\sim\crcr}}}
\def\simlt{\mathrel{\mathpalette\mathrelfun <}}
\def\simgt{\mathrel{\mathpalette\mathrelfun >}}

\def\ln {{\rm ln}}
\def\sgn {{\rm sgn}}

\def\rmb {{\rm b}}

\def\rme {{\rm e}}

\def\rmB {{\rm B}}
\def\rmC {{\rm C}}

\def\rmT {{\rm T}}

\def\bfn {{\bf n}}

\def\bfp {{\bf p}}

\def\bfv {{\bf v}}

\def\calH {{\cal H}}

\def\calO {{\cal O}}

\def\hatbfn  {{\hat\bfn}}

\def\hatbfp  {{\hat\bfp}}

\def\pdbypd#1#2{{\partial#1\over\partial#2}}

\def\eV  {{\rm \hbox{e\kern-0.14em V}}}
\def\keV {{\rm \hbox{ke\kern-0.14em V}}}
\def\MeV {{\rm \hbox{Me\kern-0.14em V}}}
\def\GeV {{\rm \hbox{Ge\kern-0.14em V}}}
\def\TeV {{\rm \hbox{Te\kern-0.14em V}}}

\begin{opening}
\title{The CMBR Spectrum}
\subtitle{A Theoretical Introduction}
\author{Albert Stebbins}
\institute{NASA/Fermilab Astrophysics Group\\
           Box 500, Batavia, IL 60510, USA}
\end{opening}

\runningtitle{The CMBR Spectrum}

\begin{document}

\FNALpptheadhighest{NASA/Fermilab Astrophysics Center}{Fermilab-Conf-97/134-A}

\section{Introduction}

	The Cosmic Microwave Background Radiation (CMBR) provides a strong
observational foundation for the standard cosmological scenario, the Big Bang
theory.  It is difficult to understand how to produce a $2.7^\circ$K blackbody
spectrum except in the context of the Big Bang scenario. The near blackbody
spectrum of the CMBR along with it's near isotropy provides compelling evidence
for a period of fairly quiescent Friedman-Robertson-Walker expansion for many
expansion time before recombination.  The past decade has seen huge advances in
the measurement of the CMBR, with COBE's definitive discovery of anisotropies
and measurement of a near perfect blackbody spectrum.  The small deviations
from isotropy have and will continue to tell us a great deal about the
inhomogeneities in our universe, and small deviations from a blackbody spectrum
can also tell us about the energetics in our universe.  Such deviations have
already been discovered in the direction of clusters of galaxies, although the
mean CMBR spectrum is, so far, indistinguishable from a blackbody spectrum.

	Here we give a introduction to the observed spectrum of the CMBR and
discuss what can be learned about it.  Particular attention will be given to
how Compton scattering can distort the spectrum of the CMBR.  This is left
toward the end though.  Unfortunately the author has no expertise in the area
of how these measurements are made but Smoot has covered this area in his
lectures.  An incomplete bibliography of relevant papers is also provided.
Some old but still highly useful reviews of the physics behind the spectra are
by Danese and De\,Zotti\cite{DaneseDeZotti77}, and Sunyaev and
Zel'dovich\cite{SZ80}. Theoretically not much has changed in this field in over
25 years.  Much of the interesting work was done by Zel'dovich and Sunyaev in
1969\cite{ZS69}.

\section{Executive Summary}

The universe today is fairly diffuse and cold, however the universe is observed
to be expanding, and in the past we may deduce that the universe was more dense
and because of $p\,dV$ work the matter in the universe would also have been
hotter. Extrapolating the expansion back to very early epochs the universe
would have been very hot and very dense and the universe must have been
expanding very rapidly in order to have grown as large as it is observed to be.
Hence the Hot Big Bang. When the matter in the universe is hot and dense the
thermal equilibration time becomes very short.  Thus we expect everything to
rapidly approach thermal equilibrium and we therefore expect the photons in the
universe to have a thermal (blackbody) spectrum at early times. It is easily
shown that expansion of the universe (or traversal through any gravitational
field) leaves a blackbody spectrum a blackbody spectrum although the
temperature may change.  This temperature change is known as the redshift and
can sometimes be thought of as either a Doppler shift or a gravitational
redshift.  Formally speaking the two may be thought of as the same phenomena
and there is often no physical sense in trying to separate them.

Thus as a first approximation we expect the photons in the universe to have a
blackbody spectrum.  The fact that the cosmic microwave background radiation
(CMBR) has nearly a blackbody spectrum is strong evidence for the Hot Big Bang
hypothesis.  There is simply not enough matter around today to thermalize so
many photons (there are $\simgt10^9$ photons for every atom) and in any case
most of the matter in our universe is much hotter that 3\,K.  The reason we
might expect a deviation from blackbody is because some of the matter in the
universe has gone out of thermal equilibrium with the photons and may either
heat or cool the photons. This can be done by non-equilibrium scattering or
absorption of existing photons or by non-equilibrium emission of new
photons. Clearly most of the matter in the universe is not today in thermal
equilibrium with the CMBR and the spectrum offers us a probe of this.  However
there are so many more CMBR photons in the universe than there are protons or
electrons that it is difficult for the matter to significantly distort the
spectrum of the CMBR.  Thus the fact that the observed CMBR spectrum is so
close to a blackbody should come as no surprise.

\section{Measures of Temperature}

The {\sl brightness} or {\sl specific intensity} of light, $I_\nu$, is defined
as the incident energy per unit area, per unit solid angle, per unit frequency,
per unit time. It may be written
\begin{equation}
I_\nu={2h\nu^3\over c^2}n_\gamma
\end{equation}
where $\nu$ is the frequency and $n_\gamma(\nu)$ is the quantum-mechanical
occupation number, i.e. the number of photons (in each polarization state) per
unit phase space volume measured in units of $h^3$. Here $h$ is Planck's
constant, and we assume the light is not (linearly) polarized so that there are
an equal number of photons in each polarization state. A {\sl blackbody} or
{\sl Planck  spectrum } has
\begin{equation}
n^{\s\rm BB}_\gamma={1\over\exp\left({h\nu\over kT}\right)-1}
\label{BlackBody}
\end{equation}
where $T$ is the the temperature.  The high-frequency ($h\nu\gg kT$) limit of
the Planck spectrum is known as {\sl Wien's law}:
\begin{equation}
I^{\s\rm W}_\nu={2h\nu^3\over c^2}\exp\left(-{h\nu\over kT}\right)
\end{equation}
while the low frequency ($h\nu\ll kT$) limit of the Planck spectrum is known as
the {\sl Rayleigh-Jeans law}:
\begin{equation}
I^{\s\rm RJ}_\nu={2\nu^2 kT\over c^2}\ .
\end{equation}
Note that the intensity is proportional to the temperature in this case.  One
may invert the Planck spectrum and characterize the intensity by the {\sl
thermodynamic temperature} or {\sl brightness temperature}:
\begin{equation}
T_\rmb={h\nu\over k\,\ln\left|1+{\displaystyle{2h\nu^3\over c^2I_\nu}}\right|}
\end{equation}
Occasionally radio astronomers may define the brightness temperature by it's
Rayleigh Jeans limit:
\begin{equation}
T^{\s\rm RJ}_\rmb={c^2I_\nu\over k\,2\nu^2}\ .
\end{equation}
In the radio region this is an excellent approximation to the thermodynamic
temperature and is simply related to the intensity, and is therefore closer to
what is actually measured.

	Here we are interested in small deviations from a blackbody spectrum,
i.e. we have some temperature $T_\gamma$ which is a good fit to $T_\rmb$ at
many frequencies, and want to express the actual spectrum in terms of small
deviations from a blackbody with this temperature, in particular in terms of
the deviations in intensity from the blackbody spectrum, $\Delta I_\nu$.  For
small deviations the deviation in brightness temperature is
\begin{equation}
\Delta T_\rmb\equiv T_\rmb-T_\gamma=
{(e^x-1)^2\over x^2 e^x}\,{c^2\Delta I_\nu\over k\,2\nu^2} \qquad
 x\equiv{h\nu\over kT_\gamma} \ .
\end{equation}
Experimentally it is often easier to measure differences rather than absolute
numbers: differences in intensity in different directions on the sky, or
between the sky and internal calibrators.  Note that in the Wien region
differences in brightness temperature are greater than in the Rayleigh-Jeans
region for the same difference in intensity.  In what follows we will tend to
plot spectral distortions in terms of differences in brightness temperature
versus the dimensionless frequency, $x$.  These ``derived'' quantities are
probably more appealing to a theorist than an observer since they are further
removed from what is actually measured.

\section{Measured Mean Spectrum of the CMBR}

\begin{figure}
\vspace{5cm}
\epsfysize=5.0cm \epsfbox[-100 144 592 450]{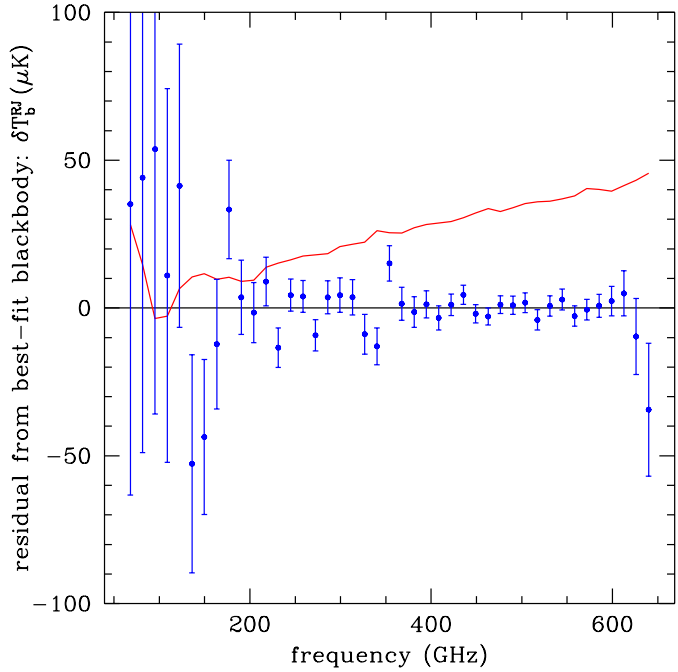}
\caption{Plotted are the residuals in Rayleigh-Jeans temperature from the best
fit blackbody as a function of frequency as stated by Fixsen {\it et
al.} (1996).  The error bars are 1-$\sigma$. The solid  line is the
subtracted Galaxy model at the Galactic poles. We see that these measurements
are running up against a fundamental limitation of Galactic contamination.} 
\label{fig:FIRAS96}
\end{figure}

\begin{figure}
\vspace{5cm}
\epsfysize=5.0cm \epsfbox[-100 144 592 450]{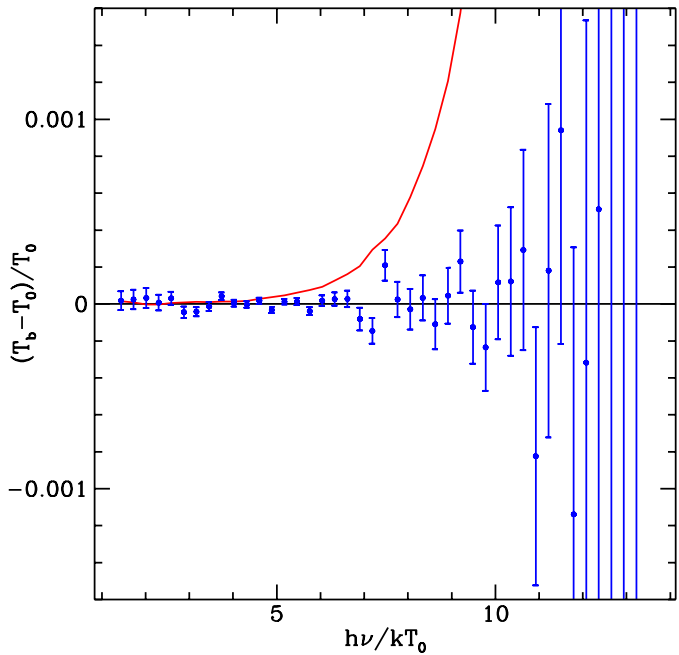}
\caption{Same as fig~1 except here we plot the fractional deviation in
brightness temperature vs. the dimensionless frequency $x={h\nu\over kT}$.  To
do this we have used the best-fit photon temperature $T_\gamma=2.728$\,K.
Plotting things in this way accentuates the deviations at high frequencies.}
\label{fig:MonopoleSpectrum}
\end{figure}

\begin{table}[htb]
\begin{center}
\caption{Listed are measurements, made over the past 15 years, of the absolute
CMBR brightness temperature at a variety of wavelengths.  The results often
include significant corrections for Galactic emission.  Millimeter wavelengths
are omitted as they have been superseded by results of FIRAS ($>68$\,GHz -
see Fixsen {\it et~al.}~1996).  The ADS code given refers to the paper where
these results are presented or reviewed and may be used to find the papers and
abstracts online at the NASA Astrophysics Data System and mirror sites: {\sl
adsabs.harvard.edu, cdsads.u-strasbg.fr, d01.mtk.nao.ac.jp}\,.  These codes are
of the form {\it year} {\it journal} {\it volume} {\it page}.}
\begin{tabular}{rrlcl}
\hline 
   Frequency     &Wavelength &$T_{\rm CMBR}$&1st Author &ADS Bibliographic \\
(GHz)\phantom{xx}&(cm)\phantom{xxx}&(Kelvin)&           &\hskip27pt   Code \\
\hline
  1.47\phantom{0}&20.4\phantom{00}&$2.26\phantom{0}^{+0.19 }_{-0.19 }$
                                           &Bensadoun  &1993ApJ...409....1B \\
 90.\phantom{000}& 0.22\phantom{0}&$2.60\phantom{0}^{+0.09 }_{-0.09 }$
                                           &Bersanelli &1989ApJ...339..632B \\
  2.0\phantom{00}&15.\phantom{000}&$2.55\phantom{0}^{+0.14 }_{-0.14 }$
                                           &   "       &1994ApJ...424..517B \\
  3.7\phantom{00}& 8.1\phantom{00}&$2.59\phantom{0}^{+0.13 }_{-0.13 }$
                                           &De Amici   &1988ApJ...329..556D \\
  3.8\phantom{00}& 7.9\phantom{00}&$2.64\phantom{0}^{+0.07 }_{-0.07 }$
                                           &   "       &1990ApJ...359..219D \\
  3.8\phantom{00}& 7.9\phantom{00}&$2.64\phantom{0}^{+0.06 }_{-0.06 }$
                                           &   "       &1991ApJ...381..341D \\
 25.\phantom{000} & 1.2\phantom{00}&$2.783^{          +0.025}_{-0.025}$
                                           &Johnson    &1987ApJ...313L...1J \\
  7.5\phantom{00}& 4.0\phantom{00}&$2.60\phantom{0}^{+0.07 }_{-0.07 }$
                                           &Kogut      &1990ApJ...355..102K \\
  1.410          &21.26\phantom{0}&$2.11\phantom{0}^{+0.38 }_{-0.38 }$
                                           &Levin      &1988ApJ...334...14L \\
  7.5\phantom{00}& 4.\phantom{000}&$2.64\phantom{0}^{+0.06 }_{-0.06 }$
                                           &   "       &1992ApJ...396....3L \\
  4.75\phantom{0}& 6.3\phantom{00}&$2.70\phantom{0}^{+0.07 }_{-0.07 }$
                                           &Mandolesi  &1986ApJ...310..561M \\
  2.5\phantom{00}&12.\phantom{000}&$2.79\phantom{0}^{+0.15 }_{-0.15 }$
                                           &Sironi     &1986ApJ...311..418S \\
  0.600          &50.\phantom{000}&$3.0\phantom{00}^{+1.2  }_{-1.2  }$
                                           &   "       &1990ApJ...357..301S \\
  2.5\phantom{00}&12.\phantom{000}&$2.50\phantom{0}^{+0.34 }_{-0.34 }$
                                           &   "       &1991ApJ...378..550S \\
  0.82\phantom{0}&36.6\phantom{00}&$2.7\phantom{00}^{+1.6  }_{-1.6  }$
                                           &   "       &\hskip40pt "        \\
  2.5\phantom{00}&12.0\phantom{00}&$2.78\phantom{0}^{+0.3  }_{-0.3  }$
                                           &Smoot      &1987ApJ...317L..45S \\
 33.0\phantom{00}& 0.909          &$2.81\phantom{0}^{+0.2  }_{-0.2  }$
                                           &   "       &\hskip40pt "        \\
  1.41\phantom{0}&21.2\phantom{00}&$2.22\phantom{0}^{+0.55 }_{-0.55 }$
                                           &   "       &\hskip40pt "        \\
  3.66\phantom{0}& 8.2\phantom{00}&$2.59\phantom{0}^{+0.14 }_{-0.14 }$
                                           &   "       &\hskip40pt "        \\
 10. \phantom{00}& 3.0\phantom{00}&$2.61\phantom{0}^{+0.06 }_{-0.06 }$
                                           &   "       &\hskip40pt "        \\
 90.\phantom{000}& 0.33\phantom{0}&$2.60\phantom{0}^{+0.10 }_{-0.10 }$
                                           &   "       &\hskip40pt "        \\
  1.4\phantom{00}&21.\phantom{000}&$2.65\phantom{0}^{+0.33 }_{-0.30 }$
                                           &Staggs     &1993PhDT.........6S \\
 10.7\phantom{00} &2.80\phantom{0}&$2.730^{          +0.014}_{-0.014}$
                                           &   "       &1996ApJ...473L...1S \\
 90.\phantom{000} &0.33\phantom{0}&$2.57\phantom{0}^{+0.12 }_{-0.12 }$
                                           &Witebsky   &1986ApJ...310..145W \\
\hline
\end{tabular}
\end{center}
\end{table}

\begin{figure}
\vspace{5cm}
\epsfysize=5.0cm \epsfbox[-100 144 592 450]{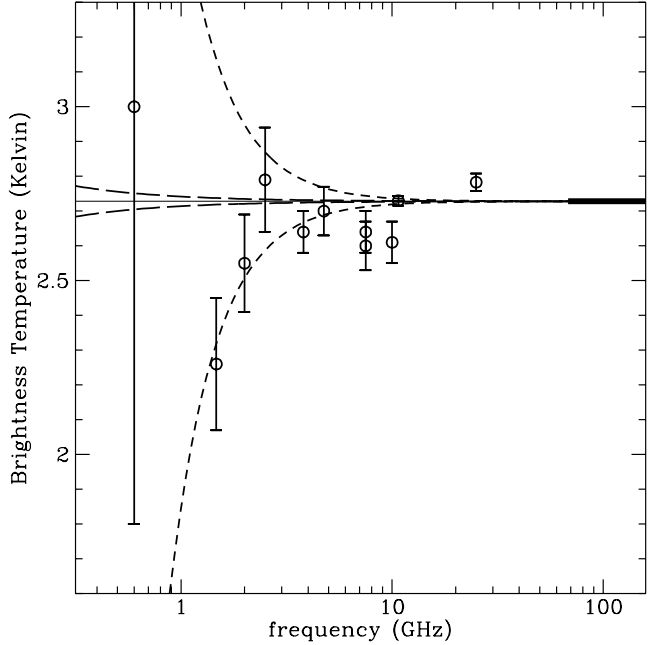}
\caption{Plotted are a selection of the low frequency measurements of the CMBR
brightness temperature listed in table 1.  From left to right the points are
from Sironi {\it et al.} (1990), Bensadoun {\it et al.} (1993), Bersanelli {\it
et al.}  (1994), Sironi\& Bonelli (1986), De\,Amici {\it et al.}  (1991),
Mandolesi {\it et al.} (1986), Kogut {\it et al.} (1990), Levin {\it et al.}
(1992), Smoot {\it et al.} (1987), Staggs (1996), Johnson \& Wilkinson (1987)
and were chosen because of the small errorbars.  The black band at the right
indicates the FIRAS data (Fixsen {\it et al.} 1996), while the horizontal
straight line represents a temperature 2.728\,K given by the FIRAS best fit
blackbody spectrum.  The long-dashed represents a chemical potential
distortions with amplitude $\mu=\pm9\times10^{-5}$ while the solid line gives
free-free distortions with amplitude $Y_{\rm ff}=\pm10^{-4}$.  These are both
idealized curves and one may expect (model dependent) corrections long-ward of
10\,GHz (see Burigana, De\,Zotti, and Danese 1995).}
\label{fig:LowFrequency}
\end{figure}

\begin{figure}
\vspace{5cm}
\epsfysize=5.0cm \epsfbox[-100 144 592 450]{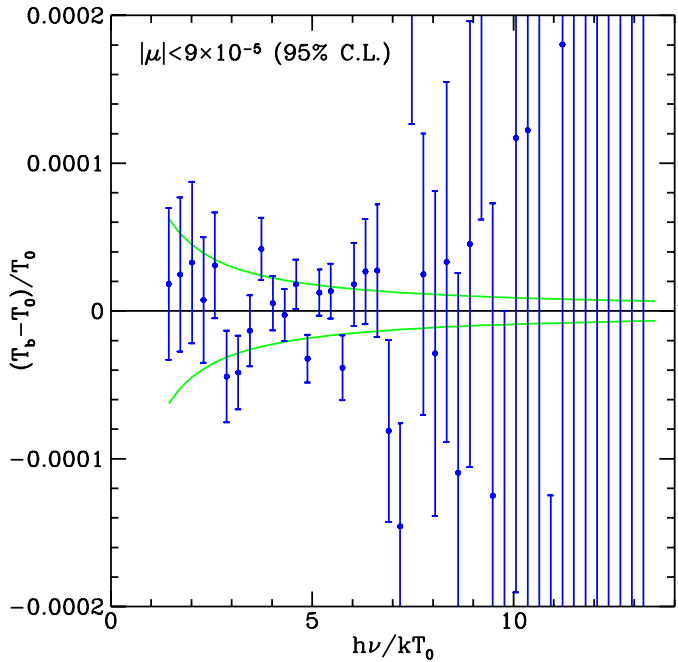}
\caption{Superimposed on the FIRAS data of fig~2 is the largest chemical
potential distortion allowed by the data (Fixsen {\it et al.} 1996): 
$\mu=\pm9\times10^{-5}$.  The falling positive curve is the far more plausible
positive chemical potential distortion and the negative rising curve is a
negative chemical potential distortion.}
\label{fig:MuDistortion}
\end{figure}

\begin{figure}
\vspace{5cm}
\epsfysize=5.0cm \epsfbox[-100 144 592 450]{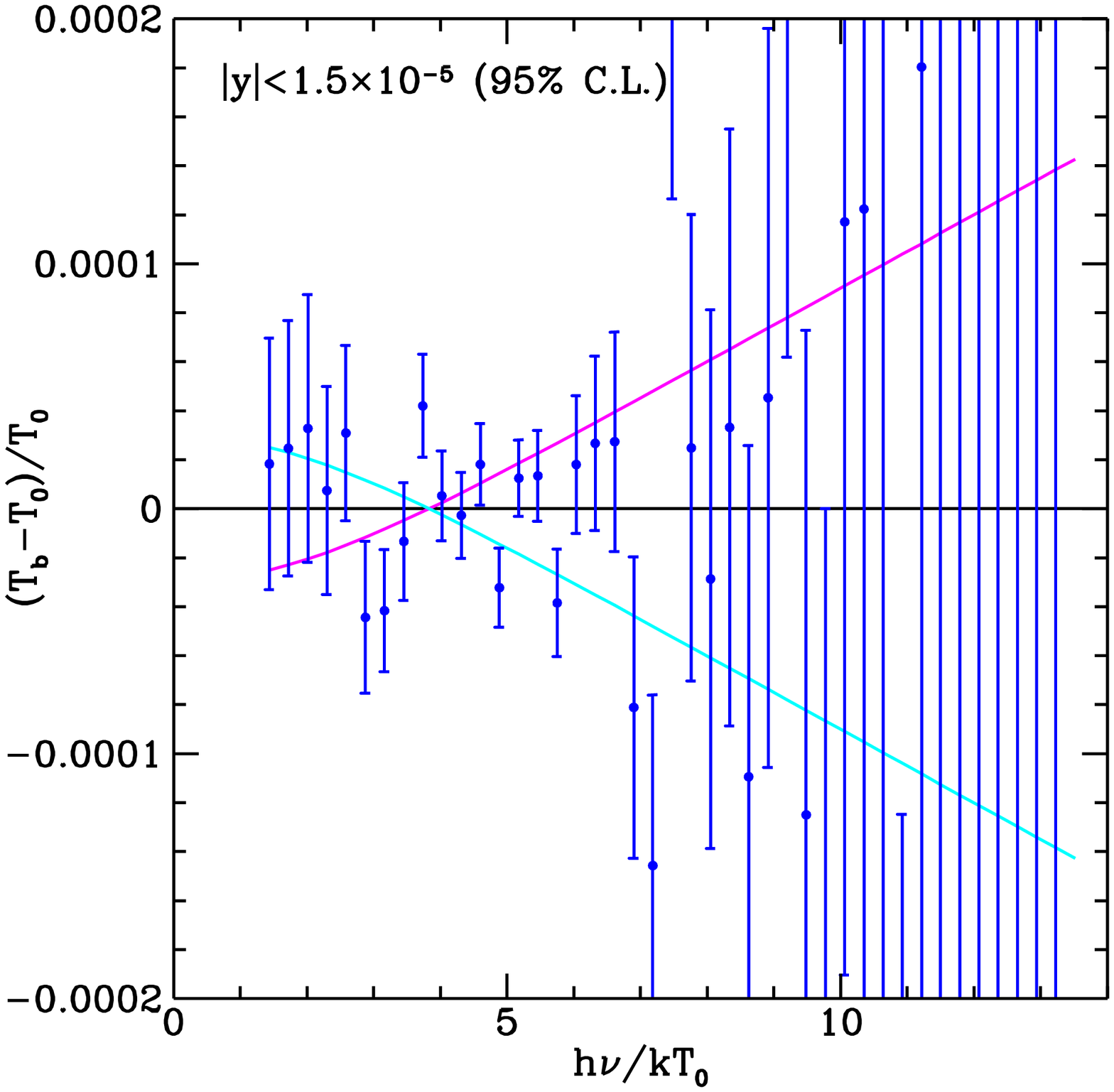}
\caption{Superimposed on the FIRAS data of fig~2 is the largest $y$-distortion
allowed by the data (Fixsen {\it et al.} 1996):  $y=\pm1.5\times10^{-5}$.  The
rising curve is the more plausible positive $y$-distortion and the falling
curve is a negative $y$-distortion.}
\label{fig:yDistortion}
\end{figure}

	Over the years there have been many measurements of the CMBR.  There
have been many claims that the spectrum deviated significantly from a
blackbody, especially in the Wien region, however recent measurements with 
FIRAS (Far-InfraRed Absolute Spectrophotometer) on the COBE satellite has shown
definitively that the spectrum is very close to a blackbody\cite{Mather90}.
Contemporary and quite competitive with with the first FIRAS measurements was a
rocket experiment \cite{Gush90} which also found a blackbody.  The most recent
FIRAS results have appeared in ref~\cite{Fixsen96} which are plotted in
figs~\ref{fig:FIRAS96}~\&~\ref{fig:MonopoleSpectrum}.  In the 2nd figure the
we have converted to a more theoretical representation by plotting versus
$x={h\nu\over kT_\gamma}$ and converting to fractional changes in temperature.
The reason that this transformation is useful is that we can predict the shape
of the deviations from blackbody in terms of $x$, which we cannot in terms of
$\nu$ since we have no {\it a priori} knowledge of $T_\gamma$.  To transform to
an $x$ variable one must decide on a fiducial temperature.  We have used the
best-fit temperature, $T_\gamma=2.728$K, taken from the most recent results of
FIRAS\cite{Fixsen96}: $T_\gamma=2.728\pm0.002$K.  Note that the uncertainty
in $T_\gamma$ is  which is larger than the error bars on most of the individual
points in the plots.  The reason that the uncertainty in $T_\rmb-T_\gamma$ can
be smaller than the uncertainty in $T_\gamma$ is that the experiment measures
the difference in brightness between a blackbody and the sky.  The $\pm0.002$K
represents the uncertainty in the temperature of the reference blackbody, while
the accuracy to which this reference is thought to be a blackbody is much
better than this.  Note that when fitting for a distortion from a blackbody one
must fit for both the amplitude of the distortion and for $T_\gamma$
simultaneously.

	While FIRAS certainly revolutionized the field, and does make obsolete
most other short wavelength measurements of the CMBR spectrum, it only looked
at the frequency range $68-640$GHz.  The bolometric techniques used by FIRAS
only work at high frequencies and therefore the spectrum at low frequency was
not touched by FIRAS.  There has been ongoing measurements of the absolute CMBR
flux in the Rayleigh-Jeans region for 30 years and we list some results from
the last 15 years in Table 1.  One can see that measurement did not stop after
COBE.  As we shall see some of the spectral distortions we are looking for are
most visible in the Rayleigh-Jeans regime.  We have selected some of the most
sensitive of measurements to plot in fig~\ref{fig:LowFrequency}.  The
uncertainties vary widely with frequency and are orders of magnitude larger
than those of FIRAS.  Several authors have noted that these low frequency
measurements tend to indicate a temperature lower than that obtained at higher
frequencies\cite{Sironi90,Bersanelli94}, suggesting that there may be a
deviation from a blackbody spectrum at low frequencies.

	Measurement of the absolute CMBR spectrum, at the present level of
sensitivity, face significant problems of Galactic contamination at both long
and short wavelengths.  Synchrotron radiation contaminates the long-wavelength
spectrum while the short wavelength region is contaminated by dust emission.
Since we cannot expect to observe the CMBR from outside of the Galaxy this is a
fundamental limitation.  Many of the results plotted here include significant
corrections for this contamination.  While there is a limit to how well one can
subtract off the Galaxy, we can look forward to improvements in Galaxy
modeling using results from anisotropy experiments which will have increasingly
better sensitivity, sky coverage, and angular resolution. While anisotropy
experiments cannot generally make absolute measurements of intensity, they can
help to map out the Galaxy.

\section{Spectral Distortions of the CMBR}

While one should not be surprised that the CMBR has close to a blackbody
spectrum, there are various mechanisms which should cause deviations from a
thermal spectrum.  Now we discuss a few of them.

\subsection{Anisotropies}

The most common way in which the CMBR spectral distortion occurs is when the
photons have a blackbody spectrum in each direction but the temperatures
characterizing these spectra are different in different directions. This
direction dependent temperature difference is called anisotropy.  The
anisotropy can either be caused by Doppler/gravitational effect or because the
gas emitting the photons really did have different temperatures.

The first anisotropy discovered was the dipole anisotropy, i.e. the temperature
varies like the cosine of the angle from some point on the celestial sphere. It
is usually attributed to the Sun moving at 371km/s.  Note that to a first
approximation, when averaging over the sky, the dipole does not contribute to
the mean spectrum of the CMBR.

One way to check that measured anisotropies are what they are supposed to be is
to measure the spectrum.  For a small anisotropy the change in flux from the
mean spectrum should be proportional to the derivative of the flux of a
blackbody with respect to temperature.  FIRAS has done just that for the
dipole\cite{Fixsen94,Fixsen96} and found just what was expected.  Most modern
anisotropy experiments use many frequency channels in order to check the
spectrum of the anisotropy, or more specifically to be able to subtract off
contamination of the measurements by other effects than the anisotropy.

\subsection{Chemical Potential Distortions}

There are three processes which are important from thermalizing the CMBR
spectrum in the early universe: {\sl Compton scattering}, {\sl double Compton
scattering}, and {\sl free-free scattering} (also known as {\sl
bremsstrahlung}). Compton scattering is a much more rapid process but since it
conserves the number of photons so it can only thermalize the energy
distribution of the photons and not the number of photons.  All of these
processes become more efficient as one goes to earlier and earlier epochs and
eventually photon non-conserving processes start to become important.

There is a epoch between $z=10^5$ and $z=2\times10^6$ during which Compton
scattering is efficient in thermalizing the energy distribution while other
processes are not capable of thermalizing the photon number. During this epoch,
if the energy-to-photon ratio is perturbed from that required for a blackbody
spectrum, the spectrum will instead approach a Bose-Einstein distribution 
\begin{equation}
n^{\s\rm BE}
={1\over\exp\left({\displaystyle{h\nu\over kT_\gamma}}+\mu\right)-1}
\end{equation}
where $T_\gamma$ and $\mu$ ( the {\sl dimensionless chemical potential}) are
determined by the total energy available and the total number of photons
available. If one starts out with a thermal distribution of photons at
temperature $T_\gamma$ and injects a fractional increase in the energy density,
${\Delta U\over U}$, without significantly increasing the number of photons one
obtains
\begin{equation}
T_\rmb\approx
T_\gamma\,\left(1-\mu\left[0.456-{1\over x}\right]\right) 
\qquad {\Delta U\over U}=0.714\mu
\qquad \mu\ll1
\end{equation}
Since one must fit the observations to both $T_\gamma$ and $\mu$ one really can
only measures the ${1\over x}$ term. Double Compton scattering and free-free
scattering become increasingly more efficient at lower frequencies and there
are usually corrections to this formula at small frequencies 
$x\ll1$\cite{Burigana91,Burigana95}. These corrections are not liable to be
important for FIRAS measurements.

This distortion to the spectrum is greatest at small $x$, however the FIRAS
measurements at high frequencies are so accurate that they yield much better
constraints on $\mu$ than does the low frequency experiments.  Comparing with
the FIRAS data one finds $|\mu|<9\times10^{-5}$ at the $2\sigma$
level\cite{Fixsen96}. We compare the maximal allowed distortion to the low \&
high frequency data in figs~\ref{fig:LowFrequency}\&\ref{fig:MuDistortion},
respectively.

Thus we find the extremely stringent constraint at a very early epoch
\begin{equation}
{\Delta U\over U}<6\times10^{-5} \qquad 10^5<z<2\times10^6
\end{equation}
Of course this is not to say that one expects large energy injection at these
epochs. 

Note that for $z>10^6$ the CMBR spectrum is not telling us much about the
energetics of the universe.  However one can use Big Bang Nucleosynthesis to
probe the total energy of the universe up to $z\sim10^{10}$.

\subsection{$y$ Distortions}

	If energy is injected into the universe after $z\sim10^4$ Compton
scattering is unable to thermalize the distribution.  The fact that we observe
very little deviation from a blackbody spectrum tells us that not much energy
could have been injected compared with the thermal energy of the CMBR.  If a
small amount of energy is injected then one may solve for the linear
perturbation from a blackbody spectrum under the action of Compton 
scattering as was done by Zel'dovich and Sunyaev\cite{ZS69}.  One finds that 
the perturbation in the photon occupation number is
\begin{equation}
\Delta n=y\,{x e^x\over(e^x-1)^2}\,\left(x\,{e^x+1\over e^x-1}-4\right)
\label{SZdistortion}
\end{equation}
where the ``$y$-parameter'' is
\begin{equation}
y=\int dt \,\sigma_\rmT\,c \,N_\rme\,{k(T_\rme-T_\gamma)\over m_\rme c^2}\ ,
\label{SZparameter}
\end{equation}
$N_\rme$ is the number density of free-electrons, and $\sigma_\rmT$ is the
Thomson cross-section.  If one could manage to cool gas below the radiation
temperature one could produce a distortion with negative $y$, but typically
this distortion is produced by ionized gas which is much hotter than the
photons.  In the early universe when the density of electrons is large even a
small heating of the gas over the photon temperature may lead to a significant
distortion.  This distortion is generally referred to as a $y$-distortion when
applied to the mean CMBR spectrum, but is usually called the Sunyaev-Zel'dovich
distortion when referring to an anisotropy in the spectrum because there is
more or less hot gas in one direction than another.  Large amounts of hot
ionized gas exist in clusters of galaxies and the ``S-Z effect'' has been
observed in the directions of several clusters.  There is no evidence for a $y$
of the mean CMBR spectrum, although with sensitive enough measurements we
should see the hot gas we know is out there.  Fixsen {\it et
al.}\cite{Fixsen96} have placed a limit of $|y|<1.5\times10^{-5}$ from the
FIRAS data.  We compare the maximal distortions to the FIRAS data in
fig~\ref{fig:yDistortion}.  Note that a positive $y$-distortion produces a
negative change in $T_\rmb$ at low frequencies and positive change in $T_\rmb$
at high frequencies, just what one expect if one was heating a fixed number of
photons.  This negative $\Delta T_\rmb$ is sometimes called the
``S-Z decrement'', for the S-Z effect was first looked for at radio
wavelengths.

\begin{figure}
\vspace{5cm}
\epsfysize=5.0cm \epsfbox[-100 144 592 450]{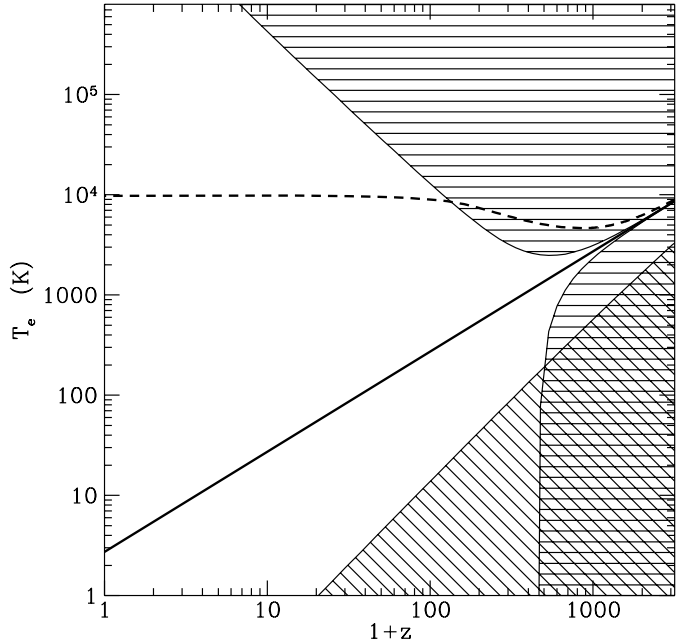}
\caption{Plotted is the constraint on the temperature of a fully ionized
universe as a function of redshift.  The horizontally hatched region is
excluded since $|y|<1.5\times10^{-5}$ while the diagonally hatched region
would be excluded if $Y_{\rm ff}<10^{-4}$.  The solid line indicates where the
gas temperature equals the photon temperature and the dashed line gives the
temperature as a function of redshift for a model where the gas is ionized by
very massive stars (VMOs - see Stebbins \& Silk 1986).  The cosmological
parameters used are $H_0=65$km/sec/Mpc, $\Omega_0=0.4$ and $\Omega_\rmb=0.10$.}
\label{fig:LoOmega}
\end{figure}

\begin{figure}
\vspace{5cm}
\epsfysize=5.0cm \epsfbox[-100 144 592 450]{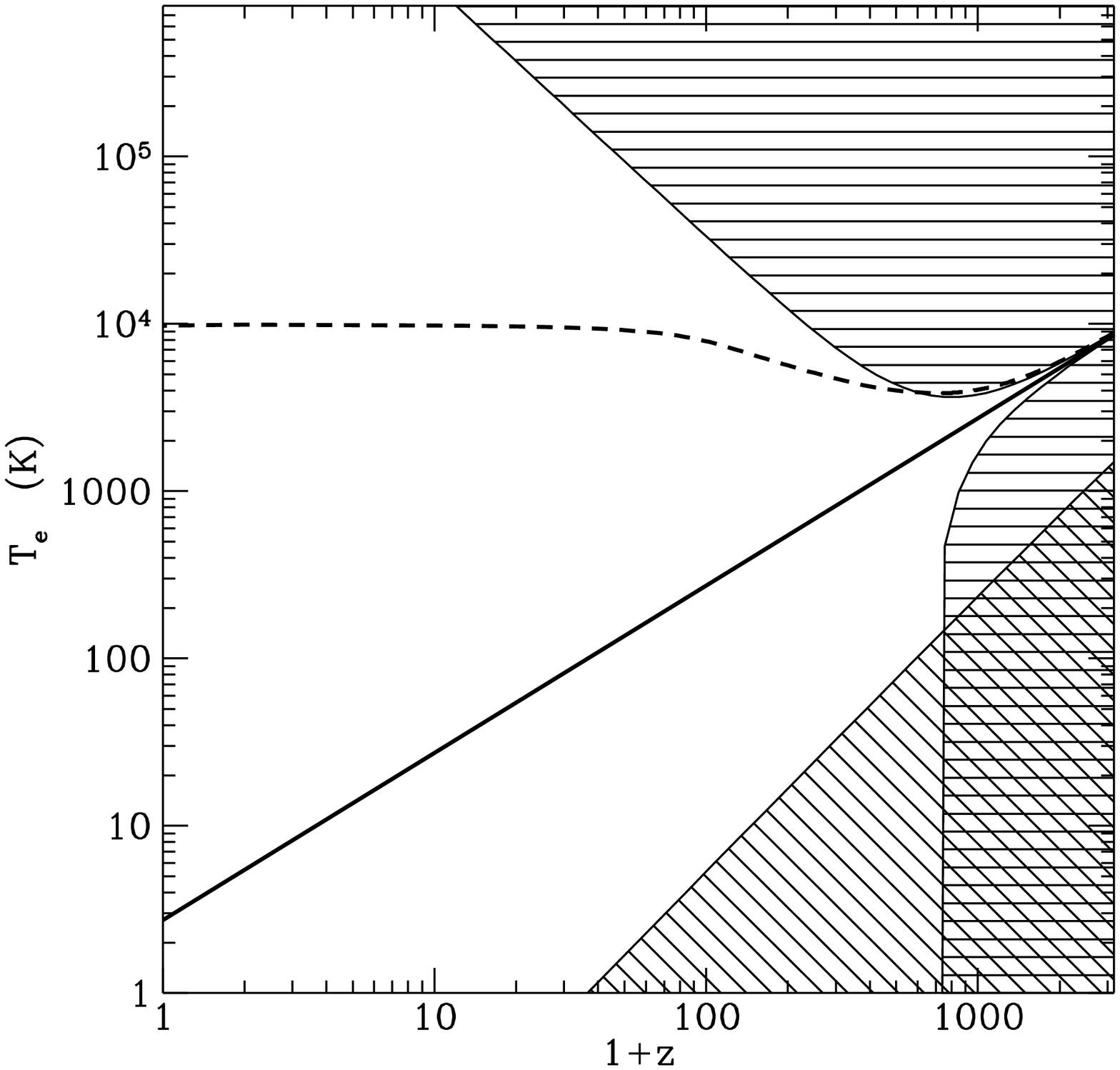}
\caption{The same as the previous figure except with $\Omega_0=1$ and
$\Omega_\rmb=0.06$.  The constraints are less severe for for larger 
$\Omega_0$ and smaller $\Omega_\rmb$.}
\label{fig:HiOmega}
\end{figure}

\subsection{Distortion from Free-Free}

	Another important process for the CMBR spectrum is free-free
scattering, which is the scattering of a free electron off of a charged nucleon
either emitting or absorbing a photon; in most cases of interest emitting
rather than absorbing.  This is the same process which produces the X-rays
observed from hot cluster gas operating at microwave and radio frequencies.
The effect of free-free scattering on the CMBR spectrum is mostly likely to be
seen at the very longest wavelengths measured. In this Rayleigh-Jeans limit the
distortion it produces may approximated by\cite{Bartlett91}
\begin{equation}
{\Delta T_\rmb\over T_\gamma}\approx{Y_{\rm ff}\over x^2}
\qquad Y_{\rm ff}=\int dt\,{T_\rme-T_\gamma \over T_\rme}\,\kappa\,dt
\end{equation}
where
\begin{equation}
\kappa\equiv{8\pi e^6h^2N_\rme^2g
             \over3m_\rme(kT_\gamma)^3\sqrt{6\pi m_\rme kT_\rme}}
\end{equation}
and the Gaunt factor, $g\sim2$ in most cases of interest.  Note the
$T_\rme^{-{1\over2}}$ factor in $\kappa$ which means that the effect is
suppressed for higher temperature gas.  The $1/x^2$ dependence on this
distortion means that it is the low frequency measurements which will constrain
it's amplitude. In fig~\ref{fig:LowFrequency} we plot free-free distortions
with $Y_{\rm ff}=\pm10^{-4}$.  We see that this size distortion is close to
what is being constrained by these measurements, although no proper statistical
analysis has been done.  This size free-free distortion would produce no
significant effect in the FIRAS data, although if one went far enough into the
Wien tail one would find large distortions from free-free emission.

	In figs~\ref{fig:LoOmega}\&\ref{fig:HiOmega} we have used constraints
on $y$ and $Y_{\rm ff}$ to put constraints on the temperature and epoch of a
reionized universe, assuming presently favored cosmological parameters. We see
that it really the $y$-distortion which is most important, the free-distortion
only being detectable on the off chance that the gas was ionized and cold.
Even though the limits on $y$ are quite small we see that there is not too much
of a constraint of ionization after $z\sim100$.  The constraints could be made
stronger if one assumes a larger baryon density or a smaller total density.

\section{Physical Processes}

	Now we will take a closer look at the physical processes which could
cause a distortion of the spectrum.  Here we will only discuss Compton
scattering although free-free emission and double-Compton deserve an equally
thorough treatment.

\subsection{Compton Scattering}

\subsubsection{Collisional Boltzmann Equation}

	One may describe the state of the primeval gas of photons and
electrons in terms of the the density of particles in phase space, i.e. 
momentum and position space.  Here we are not interested in the polarization
state of electrons and photons so we average over the two polarization states
\footnote{Compton scattering in an inhomogeneous medium will produce
some polarization of the photons, which can be measured, and also effects the
anisotropy at the several percent level.  See Melchiorri and Vittorio this
volume.}
It is convenient to measure the phase space density in units of $h=2\pi\hbar$
which gives the the quantum mechanical occupation number, $n_\gamma$ and
$n_\rme$ for photons and  electrons, respectively.  The evolution of $n_\gamma$
can be described by the collisional Boltzmann equation which has the form 
\begin{equation}
{D n_\gamma(\bfp_\gamma)\over Dt}=C(\bfp_\gamma)
\end{equation}
where $C(\bfp_\gamma)$ is the scattering term which describes the interactions
with other particles.  Here ${D\over Dt}$ is a convective derivative along the
photon's trajectory in phase space, while the right-hand-side gives the
collision integral.  If there were no collisions then the Boltzmann equation
states that the occupation number remains constant along photon trajectories.
\footnote{This is true for a phase space defined by a position, $x^i$, and it's
canonically conjugate momentum, $p_i$, $n(\bfp)$ measures the particle density
with volume measure: $d^3x^i d^3p_i$.  In general relativity there is both the
covariant momentum, $p_i$, and contravariant momentum, $p^i$.  If one measures
the density of particles per unit $d^3x^i d^3p^i$ the Boltzmann equation as
expressed above does not apply!}
Included in this convective derivative are the all the effects of gravity on
the photons, which include many of the effects which produce anisotropy.  We
will not discuss these effects further as they are covered extensively in
Bunn's lectures.

	The collision integral for Compton scattering of unpolarized particles
after averaging over the polarization state if scattered particles is of the 
form\footnote{This form is determined by the {\it principle of detailed
balance} which results from the time-reversal symmetry of the S-matrix (or
classical or quantum mechanics)\cite{Kinetics}.}
\begin{eqnarray}
&&\hskip-25pt C^\rmC(\bfp_\gamma)={2\over(2\pi\hbar)^3}
\int d^3\bfp_\rme\int d^2\hatbfn'\,
              c\,(1-\hatbfn\ccdot\vec{\beta})\,{d^2\sigma\over d^2\hatbfn'} \cr
&&\hskip30pt\phantom{\Biggl|}
  \times\Bigl[(1-n_\rme  (\bfp_\rme   ))\,n_\rme  (\bfp_\rme'  )\,
              (1+n_\gamma(\bfp_\gamma ))\,n_\gamma(\bfp_\gamma')          \cr
&&\hskip45pt -(1-n_\rme  (\bfp_\rme'  ))\,n_\rme  (\bfp_\rme   )\,
              (1+n_\gamma(\bfp_\gamma'))\,n_\gamma(\bfp_\gamma )\Bigr]
\label{CollisionIntegral}
\end{eqnarray}
where we have (or will) use the notation
\begin{eqnarray}
&&\bfp_\gamma ={\epsilon \over c}\hatbfn    \qquad
  \bfp_\gamma'={\epsilon'\over c}\hatbfn'   \qquad
     |\hatbfn|=|\hatbfn'|=1                 \qquad
             E=\sqrt{(m_\rme c^2)^2+(c\bfp_\rme)^2}  \cr
&&   \bfp_\rme=(m_\rme c)\gamma\vec{\beta} \qquad 
         \beta=|\vec{\beta}|                \qquad
        \gamma={1\over\sqrt{1-\beta^2}}     \ .
\end{eqnarray}
In eq~\ref{CollisionIntegral} the values of $\bfp_\rme'$ and $\epsilon'$ is
determined by energy-momentum conservation.  The 2nd term in square brackets
describes the scattering
$\bfp_\gamma+\bfp_\rme\rightarrow\bfp_\gamma'+\bfp_\rme'$ while the 1st term 
results from the inverse process,
$\bfp_\gamma'+\bfp_\rme'\rightarrow\bfp_\gamma+\bfp_\rme$.  The $1+n_\gamma$
factor represents the increased scattering rate due to the stimulated emission
of the bosonic photons, while the $1-n_\gamma$ is the Pauli blocking factor
giving the exclusion principle for fermionic electrons. The factor of $2$ in
the prefactor counts the two polarization states of the incoming electrons.
The factor $c(1-\hatbfn\ccdot\vec{\beta})$ in eq~\ref{CollisionIntegral} is
a measure of the relative velocity between the ingoing electron and
photon.\footnote{This relative velocity factor is really determined by the 
definition of the cross-section.  The factor is equal to 
$\sqrt{|\bfv_1-\bfv_2|^2-{1\over c^2}|\bfv_1\times\bfv_2|^2}$ which reduces to
the above expression  when one of the particles is massless.  If both incoming
particles are non-relativistic then it reduces to the ``usual''
definition of relative-velocity: $|\bfv_1-\bfv_2|$.}
Of course, ${d^2\sigma\over d^2\hatbfn'}$ is the differential Compton
cross-section\footnote{The outgoing particle momentum $\bfp_\rme$ and
$\bfp_\gamma$ are described by six numbers however four are fixed by energy and
momentum conservation.  The differential cross-section is a function of the
remaining two parameters, which in this case we have taken to be the outgoing
photon direction, $\hatbfn'$. Any two parameters would do!}

	Note that this form of the collision integral guarantees that a thermal
distribution is a fixed point.  Substituting a Fermi-Dirac distribution for the
electrons and a Bose-Einstein distribution for the photon, i.e.
\begin{equation}
n_\rme  (E       )={1\over\exp({E\over k_\rmB T}+\mu_\rme)+1} \qquad
n_\gamma(\epsilon)={1\over\exp({\epsilon\over k_\rmB T}+\mu_\gamma)+1}
\label{BoseEinstein}
\end{equation}
will cause the integrand of the collision integral to zero so long as the
temperature is same for both.  Here $\mu_\rme$, and $\mu_\gamma$ are
(dimensionless) chemical potentials given by the total electron and photon
density, each of which is conserved by Compton scattering.  We expect such a
thermal distribution to be a stable fixed point since it is the highest entropy
state and entropy increases according to Boltzmann's H-theorem\cite{Kinetics}.
In the contexts we are interested in the density of electrons is sufficiently
low that $\mu_\rme\gg1$ and Fermi-blocking is unimportant so we may set
$1-n_\rme\rightarrow1$.  In this limit the equilibrium distribution for the
electrons becomes a simple Boltzmann distribution, i.e.,
\begin{equation}
n_\rme(E)=\exp(-{E\over k_\rmB T}-\mu_\rme)\ .
\label{BoltzmannDistribution}
\end{equation}
Henceforth we will ignore Fermi-blocking.

	We are not really interested in the scattered electrons, so we may
``integrate out'' the electron distribution function.  The idea is that we
know the electron distribution function {\it a priori} - which is often is a
true since Coulomb scattering is usually very effective in thermalizing the
electron momenta.  Thus we may rewrite the collision integral as
\begin{eqnarray}
&&C^\rmC(\bfp_\gamma)
=\int d^2\hatbfp_\gamma'\,\Biggl[{\epsilon^2\over{\epsilon'}^2}
        S(\bfp_\gamma',\bfp_\gamma )\,(1+n_\gamma(\bfp_\gamma ))\,
                                              n_\gamma(\bfp_\gamma')        \cr
&&\hskip95pt
       -S(\bfp_\gamma ,\bfp_\gamma')\,(1+n_\gamma(\bfp_\gamma'))\,
                                              n_\gamma(\bfp_\gamma )\Biggr]
\end{eqnarray}
where
\begin{equation}
S(\bfp_\gamma,\bfp'_\gamma)
={2\over(2\pi\hbar)^3}\int d^3\bfp_\rme n_\rme(\bfp_\rme)\,
  (1-\vec{\beta}\ccdot\hatbfn)
  {d^2\sigma\over d^2\hatbfn'}(\bfp_\rme,\bfp_\gamma\,\hatbfn')\,
  {\delta(\epsilon'-\epsilon(1+\bar{\Delta}))\over{\epsilon'}^2}
\end{equation}
and $\bar{\Delta}(\bfp_\rme,\bfp_\gamma,\hatbfn')$ gives the fractional
change in the energy determined by energy-momentum conservation, i.e. is the 
solution to the equation
\begin{eqnarray}
&&\hskip-20pt\sqrt{(m_\rme c^2)^2+|c\bfp_\rme|^2}+c|\bfp_\gamma|            \cr
&&=\sqrt{(m_\rme c^2)^2
               +|c\bfp_\rme+c\bfp_\gamma-\epsilon\,(1+\bar{\Delta})\hatbfn'|^2}
 +c\,|\bfp_\rme|\,(1+\bar{\Delta}) .
\end{eqnarray}
A unique solution always exists with $\bar{\Delta}\in[-1,\infty)$.

	We know that the CMBR is very nearly isotropic today, and it is
reasonable to assume that the background radiation was alway isotropic.
Since we are interested in changes in the spectrum and not anisotropy we may 
also average the collision integral over $\hatbfn$ to find the mean change in
the spectrum.  Performing the two averages $\hatbfn$ and $\hatbfn'$ the
collision integral becomes
\begin{eqnarray}
&&\hskip-13pt C^\rmC(\epsilon,\Delta)
=\int d\Delta\,
\Biggl[{1\over(1+\Delta)^3}\overline{S}({\epsilon\over1+\Delta},\Delta)
         \,(1+n_\gamma(\epsilon))\,n_\gamma({\epsilon\over1+\Delta})        \cr
&&\hskip110pt -          \overline{S}(\epsilon,\Delta)
         \,(1+n_\gamma(\epsilon(1+\Delta)))\,n_\gamma(\epsilon)\Biggr]\ .
\label{EnergyCollisionIntegral}
\end{eqnarray}
where
\begin{equation}
\overline{S}(\epsilon,\Delta)=
{\epsilon^3(1+\Delta)^2\over4\pi}\int d^2\hatbfn\int d^2\hatbfn'
           \,S({\epsilon\over c}\hatbfn,{\epsilon\over c}(1+\Delta)\hatbfn')\ .
\label{EnergyCollisionKernel}
\end{equation}
To obtain eq~\ref{EnergyCollisionIntegral} we have used a little trick of
changing the variable of integration for inverse scattering from $\Delta$ to
${1\over1+\Delta}-1$, and renaming this new dummy variable $\Delta$.  If one
looks closely at eq~\ref{EnergyCollisionIntegral} one can see that the total
photon number is preserved by scattering no matter what the form of
$\overline{S}(\epsilon,\Delta)$. 

\subsubsection{Fokker-Planck Equation}

	One important property of cosmological Compton scattering is that, at
the low redshifts we are interested in, the background radiation photons have
much lower (total) energy and are moving much faster than the electron they are
scattering off of.  One is bouncing a very light object (the photon) off of a
much more slowly moving heavy object (the electron) and energy and momentum
conservation dictates that that energy of the light object is nearly unchanged
by the scattering (consider bouncing a ping-pong ball off of a bowling ball).
The electrons are not infinitely massive nor are they completely stationary so
that the photon energy will change slightly in each collision.  All this will
be reflected in the fact that $\overline{S}(\epsilon,\Delta)$ when considered
as a function of $\Delta$ will be a very narrow function sharply peaked around
$\Delta=0$ with width much less than unity.  In contrast the
$\Delta$-dependence of $n_\gamma(\epsilon(1+\Delta))$ and
$\overline{S}({\epsilon\over1+\Delta},)$ is a much smoother function in the
sense that they do not vary much over the region in $\Delta$ where
$\overline{S}(,\Delta)$ is significantly non-zero.  Thus it should be a good
approximation to Taylor expand the integrand of
eq~\ref{EnergyCollisionIntegral} in $\Delta$ about $\Delta=0$, but excluding
the rapid dependence through the 2nd argument of $\overline{S}$ and truncating
at a given order.  This is a kind of Fokker-Planck equation\footnote{Fokker and
Planck actually considered the case where the momentum is only slightly changed
in each scattering and proposed Taylor expanding to 2nd order in the small
change in momentum.  For Compton scattering the direction of the photon will
change significantly so the momentum change is not small, but the energy change
is, and expanding in the small fractional energy change, $\Delta$, is an
obvious generalization.  It is useful to consider expanding to higher order
than 2nd.}.  If we expand to 2nd order in $\Delta$ the Boltzmann equation
becomes a partial differential equation (see eq~8 of ref~\cite{Barbosa82})
\begin{equation}
{D n_\gamma\over D\tau}
={1\over\epsilon^2}\pdbypd{}{\epsilon}\,
\left[\epsilon^3\left( {1\over2}\epsilon\,\overline{\overline{\Delta^2}}\,
                                          \pdbypd{n_\gamma}{\epsilon}
                   +\left(-                 \overline{\overline{\Delta}}
                          +    2            \overline{\overline{\Delta^2}}
                          +{1\over2}\epsilon\,\overline{\overline{{\Delta^2}'}}
                          \right)\,(1+n_\gamma)\,n_\gamma\right)\right]
\label{EnergyFokkerPlanckTwo}
\end{equation}
where
\begin{equation}
\overline{\overline{ \Delta^n  }}={1\over N_\rme \sigma_\rmT}
\int_{-1}^\infty d\Delta\,\Delta^n\,\overline{S} (\epsilon,\Delta)
\qquad
  \overline{\overline{{\Delta^n}'}}={1\over N_\rme \sigma_\rmT}
\int_{-1}^\infty d\Delta\,\Delta^n\,
                    \pdbypd{\overline{S}(\epsilon,\Delta)}{\epsilon}
\end{equation}
and we have used the electron density, $N_\rme$, introduced the Thomson
cross-section\footnote{Thomson scattering is the non-relativistic and classical
limit of Compton scattering as first described by J.J.~Thomson.},
$\sigma_\rmT$, and defined the Thomson optical depth, $\tau$:
\begin{equation}
N_\rme={2\over(2\pi\hbar)^3}\int d^3\bfp_\rme\,n_\rme(\bfp_\rme)\qquad
\sigma_\rmT={8\pi\over3}\left({e^2\over m_\rme c^2}\right)^2    \qquad
d\tau=N_\rme c\,\sigma_\rmT dt\ .
\end{equation}
This optical depth gives the expected number of Compton scatterings of low
energy photons off of non-relativistic electrons.

	The form of the equation
is reminiscent of a diffusion equation which is good description of the
physics, the small changes in photon energy at each scattering causes the
photons to diffuse in energy space. The $\pdbypd{n_\gamma}{\epsilon}$ term
causes a net drift toward increasing energies while the $(1+n_\gamma)n_\gamma$
will cause a net drift to lower energies (if it's coefficient is positive).  We
expect these drifts and diffusion to sum to zero in thermal equilibrium, i.e.
when $n_\gamma$ has a Bose-Einstein distribution (eq~\ref{BoseEinstein}), the
electrons have a Boltzmann distribution (eq~\ref{BoltzmannDistribution}),
and the two share a common temperature.  This consideration alone suggest that
for a thermal electron distribution with temperature $T_\rme$ that we should
expect
\begin{equation}
{-2          \overline{\overline{\Delta}}
 +4          \overline{\overline{\Delta^2}}
 + \epsilon\,\overline{\overline{{\Delta^2}'}}\over
             \overline{\overline{\Delta^2}}        }
={\epsilon\over kT_\rme}\ .
\end{equation}
Another feature of eq~\ref{EnergyFokkerPlanckTwo} is the differential operator
${1\over\epsilon^2}\pdbypd{}{\epsilon}$ in front, which  guarantees
conservation of photon number.  This will persist to all order in the
$\Delta$-expansion. In fact one can pretty much guess the 2nd order 
Fokker-Planck equation without knowing much about the Compton cross-section.
We will take a more constructive approach below.

\subsubsection{Compton Cross-Section}

To compute the Compton collision integral, or it's Fokker-Planck approximations
one needs to use the Compton cross-section.  The relativistic (Klein-Nishina)
differential Compton cross-section in an arbitrary rest-frame
is\cite{Barbosa82}
\begin{eqnarray}
&&\hskip-10pt
{d^2\sigma\over d^2\hatbfn'}={3\sigma_\rmT\over16\pi}\,
{1-\beta^2\over
 [1-\hatbfn'\ccdot\vec{\beta}+\alpha\gamma^{-1}(1-\hatbfn\ccdot\hatbfn')]^2}\cr
&&\hskip50pt
  \times\Biggl[1+\left(1-{(1-\beta^2)(1-\hatbfn \ccdot\hatbfn')
                       \over(1-\hatbfn \ccdot\vec{\beta})\,
                            (1-\hatbfn'\ccdot\vec{\beta})  }\right)^2       \cr
&&\hskip75pt    +{\alpha^2(1-\beta^2)\,(1-\hatbfn\ccdot\hatbfn')^2\over
                                 (1-\hatbfn'\ccdot\vec{\beta})
               \,[1-\hatbfn'\ccdot\vec{\beta}
                  +\alpha\gamma^{-1}(1-\hatbfn\ccdot\hatbfn')]}\Biggr]
\label{ComptonCrossSection}
\end{eqnarray}
where
\begin{equation}
\alpha={\epsilon\over m_\rme c^2} \ .
\end{equation}
For many astrophysical applications, and especially those related the the CMBR
there are two small numbers which enter this cross-section.  Firstly $\alpha$
is very small for the microwave photons we see observe today, roughly
$10^{-9}$.  Clearly as we go to higher redshifts the background photons become
more energetic, but $\alpha$ remains small in the redshift range relevant to
the CMBR spectrum $z\simlt 10^7$.  The 2nd small number is $\beta$ since we
are almost always interested in non-relativistic electrons.  If one is
interested in a thermal electron velocity distribution then a small $\beta$
expansion is equivalent to a small ${kT_\rme\over m_\rme c^2}$ expansion.
In most applications the $\alpha\ll{kT_\rme\over m_\rme c^2}$ so we will
concentrate more on the higher order terms in $T_\rme$ and not $\alpha$.

\begin{figure}
\vspace{1.5cm}
\epsfysize=5.0cm \epsfbox[30 170 592 400]{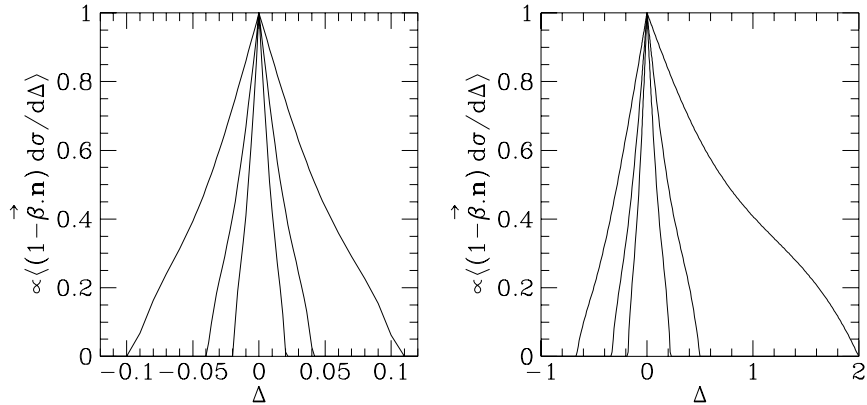}
\caption{Plotted is the distribution of fractional energy changes, $\Delta$,
experienced by low energy photons scattering off of an isotropic distribution
of electrons with velocity $\beta c$.  The left panel shows the distribution 
for $\beta=0.01$, 0.02, and 0.05; while the right panel shows the distribution
for $\beta=0.1$, 0.2, and 0.5.  For graphical clarity we have adjusted the
heights of the curves to have unit amplitude at $\Delta=0$.  The maximum and
minimum values for $\Delta$ are dictated by energy and momentum conservation.
The distribution is narrow and symmetric for small $\beta$ and becomes wider
and more skew for larger velocity electrons.  This positive skewness gives the
heating of the photons by the electrons. The Fokker-Planck equation
approximates the photon distribution function by the first few terms in it's
Taylor series about $\Delta=0$ when convolving with these distributions.  This
is liable to be a good approximation for scattering off of low velocity
electrons since the $\Delta$-distribution is sharply peaked around $\Delta=0$.}
\label{fig:DeltaDistribution}
\end{figure}

	To proceed it is probably easiest to follow the methodology of Barbosa
\cite{Barbosa82}, where one expands the cross-section in $\alpha$ but not
$\beta$.  For a thermal electron distribution one can compute the moments,
$\overline{\overline{\Delta^n}}$, in the Fokker-Planck expansion analytically,
and only at the end one should Taylor expand the result in $T_\rme$ about
$T_\rme=0$.  One may rewrite  eq~\ref{EnergyCollisionKernel} as
\begin{equation}
\overline{S}(\epsilon,\Delta)=N_\rme\,
\left\langle(1-\vec{\beta}\ccdot\hatbfn)\,{d\sigma\over d\Delta}
                        \right\rangle_{\hatbfn,\hatbfn'}
\end{equation}
where ${d\sigma\over d\Delta}$ gives the differential cross-section wrt to the
fractional change in photon energy.  So for example, expanding everything to
zeroth order in $\alpha$ (which we denote by the superscript ${}^{\s(0)}$)
we find
\begin{equation}
\left\langle(1-\vec{\beta}\ccdot\hatbfn)\,{d\sigma^{\s(0)}\over d\Delta}
                             \right\rangle_{\hatbfn,\hatbfn'}
=\sigma_\rmT\,\overline{F}(\Delta,\beta\,\sgn(\Delta))
\end{equation}
where
\begin{eqnarray}
&&\overline{F}(\Delta,b)=\sgn(\Delta)\times\calH(1-{(1-b)\,\Delta\over2b})  \cr
&&\hskip20pt\times\Biggl[
 {3(1-b^2)^2(3-b^2)(2+\Delta)\over16b^6}\,\ln{(1-b)(1+\Delta)\over1+b}      \cr
&&\hskip35pt+{3(1-b^2)(2b-(1-b)\Delta)\over32b^6(1+\Delta)}
             \Bigl(4(3-3b^2+b^4)                                            \cr
&&\hskip170pt      +2(6+b-6b^2-b^3+2b^4)\Delta                              \cr
&&\hskip170pt      +(1-b^2)(1+b)\Delta^2\Bigr)\Biggr]\ ,
\end{eqnarray}
and $\calH()$ is the Lorentz-Heaviside function which is unity for positive
argument and zero otherwise.  We see that this function is only non-zero for
\begin{equation}
\Delta\in\left[-{2\beta\over1+\beta},{2\beta\over1-\beta}\right]
\end{equation}
and, as promised, for small $\beta$ is sharply peaked around $\Delta=0$.  We
plot this function for various values of $\beta$ in
fig~\ref{fig:DeltaDistribution}.

\subsubsection{Moments of $\Delta$}

	With this general expression for ${d\sigma^{\s(0)}\over d\Delta}$ given
above one can compute, to 0th order in $\alpha$, the $\Delta$-moments which are
the coefficients in the Fokker-Planck equation (some of these may be found in
ref~\cite{Barbosa82}:
\begin{eqnarray}
&&\hskip-24pt\overline{\overline{\Delta^0}}^{\s(0)}\hskip-6pt
   =1\phantom{\biggl|}                                                      \cr
&&\hskip-24pt\overline{\overline{\Delta^1}}^{\s(0)}\hskip-6pt
   ={ 4\over  3}\overline{\gamma^2\beta^2}              \hskip65pt
   =    4       \left({kT_\rme\over m_\rme c^2}\right)
    +  10       \left({kT_\rme\over m_\rme c^2}\right)^2
    +\calO\left[\left({kT_\rme\over m_\rme c^2}\right)^3\right]           \cr
&&\hskip-24pt\overline{\overline{\Delta^2}}^{\s(0)}\hskip-6pt
   ={ 2\over 15}\overline{\gamma^4\beta^2( 5+16\beta^2)}\hskip10pt
   =   2        \left({kT_\rme\over m_\rme c^2}\right)
    + 47        \left({kT_\rme\over m_\rme c^2}\right)^2
    +\calO\left[\left({kT_\rme\over m_\rme c^2}\right)^3\right]           \cr
&&\hskip-24pt\overline{\overline{\Delta^3}}^{\s(0)}\hskip-6pt
   ={ 4\over 25}\overline{\gamma^6\beta^4(21+23\beta^2)}\hskip58pt
   ={ 252\over5}\left({kT_\rme\over m_\rme c^2}\right)^2
    +\calO\left[\left({kT_\rme\over m_\rme c^2}\right)^3\right]           \cr
&&\hskip-24pt\overline{\overline{\Delta^4}}^{\s(0)}\hskip-6pt
   ={ 4\over525}\overline{\gamma^8\beta^4(147+1554\beta^2+859\beta^4)}
   ={84\over  5}\left({kT_\rme\over m_\rme c^2}\right)^2
    +\calO\left[\left({kT_\rme\over m_\rme c^2}\right)^3\right]\ .        \cr
&&
\label{Moments}
\end{eqnarray}
and we also find that $\overline{\overline{{\Delta^n}'}}^{\s(0)}=0$ since
${d\sigma^{\s(0)}\over d\Delta}$ has no dependence on $\epsilon$.  The fact
that $\overline{\overline{\Delta^0}}^{\s(0)}\hskip-5pt=1$ tells us that, to 0th
order in $\alpha$ and all orders in $\beta$ the scattering rate per unit volume
is $c N_\rme\sigma_\rmT$.\footnote{The total (Klein-Nishina) cross-section
starts to fall below the Thomson cross-section when the center-of-mass photon
energy rises to close to $m_\rme c^2$, i.e. when $\gamma\alpha\simgt1$.  A
careful look at eq~\ref{ComptonCrossSection} will show that by setting
$\alpha=0$ in this equation we are ignoring terms of order $\alpha\gamma$.  For
microwave photons this approximation should be good for computing the total
cross-section as long as $\gamma\simlt10^9$ i.e. for anything less energetic
than $\sim500\TeV$ electrons.  In contrast to compute the small effects on the
spectrum from Compton scattering one should include 1st order terms in $\alpha$
whenever $\alpha\simgt\beta^2$.}  The coefficients in the Fokker-Planck
equations are determined by the average of the electron velocities indicated,
and these expressions hold whether or not the electrons are in thermal
equilibrium.  For a thermal distribution these velocity moments can be computed
exactly in terms of modified Bessel functions\cite{Barbosa82}, however we have
found it convenient to expand these functions to the appropriate order in
temperature. It seems that a Taylor series to a given order in $\Delta$ is less
accurate than the same order Taylor series expansion in $T_\rme$.  To keep
track of the various terms in the expansion let us devise the notation 
\begin{equation}
\calO(n,m)=\calO\left(\left({ kT_\rme\over m_\rme c^2}\right)^n
                      \left({\epsilon\over m_\rme c^2}\right)^m\right)
\label{Order}
\end{equation}
There are no terms $\sim\calO(0,0)$.  One finds that
\begin{equation}
\overline{\overline{\Delta^{2n-1}}}^{\s(m)}\sim
\overline{\overline{\Delta^{2n  }}}^{\s(m)}\sim\calO(n,m)\ .
\end{equation}
so to include all the terms of order $\sim\calO(n,m)$ in one must make a
Fokker-Planck expansion to order $2n$ in $\Delta$.  It is probably not
worthwhile to go to high order in these expansions, since one can circumvent
this expansion by doing the collision integral.  Nevertheless the first few
terms give useful analytical expressions.

\subsubsection{The Kompaneets Equation and Relativistic Corrections}

	The lowest order non-zero Fokker-Planck equation, given by the
expansion of eq~\ref{EnergyFokkerPlanckTwo}, is the Kompaneets
equation
\begin{eqnarray}
&&\hskip-15pt\pdbypd{n_\gamma}{\tau}
={1\over \epsilon^2}\pdbypd{}{\epsilon}\,
 \left[\epsilon^3\left({kT_\rme\over m_\rme c^2}\,
                       \epsilon\,\pdbypd{n_\gamma}{\epsilon}
                       +{\epsilon\over m_\rme c^2}\,
                        (1+n_\gamma)\,n_\gamma\right)\right]                \cr
&&\hskip95pt\bigl|    \hskip90pt\bigl|  \cr
&&\hskip80pt\calO(1,0)\hskip60pt\calO(0,1)
\end{eqnarray}
where the order of the two terms are indicated.  This equation was first
published by Kompaneets\cite{Kompaneets57} in 1957 and probably developed
earlier as part of the Soviet thermonuclear weapons  program.  For hotter gas
one can add terms $\calO(2,0)$ which yields an extended Kompaneets
equation\cite{Stebbins97}
\begin{eqnarray}
&&\pdbypd{n_\gamma}{\tau}
={1\over \epsilon^2}\pdbypd{}{\epsilon}\,
\Biggl[\epsilon^3\Bigg({kT_\rme\over m_\rme c^2}\,
 \left(1+{5\over2}{kT_\rme\over m_\rme c^2}\right)\,
                                      \epsilon\,\pdbypd{n_\gamma}{\epsilon} \cr
&&\hskip85pt
+{7\over10}\left({kT_\rme\over m_\rme c^2}\right)^2\,
         \left( 6\epsilon^2\,\pdbypd{^2n_\gamma}{\epsilon^2}
               + \epsilon^3\,\pdbypd{^3n_\gamma}{\epsilon^3}\right)         \cr
&&\hskip85pt
+{\epsilon\over m_\rme c^2}\,(1+n_\gamma)\,n_\gamma\Biggr)\Biggr]
\label{ExtendedKompaneets}
\end{eqnarray}
Further terms in this expansion will be derived in ref~\cite{Stebbins97}
although it is not clear how useful they will be since extensive numerical work
has been done with the more accurate collision integral (e.g.
ref~\cite{Rephaeli95}).

\subsubsection{The Generalized Sunyaev-Zel'dovich Effect}

\begin{figure}
\vspace{5cm}
\epsfysize=5.0cm \epsfbox[-100 144 592 450]{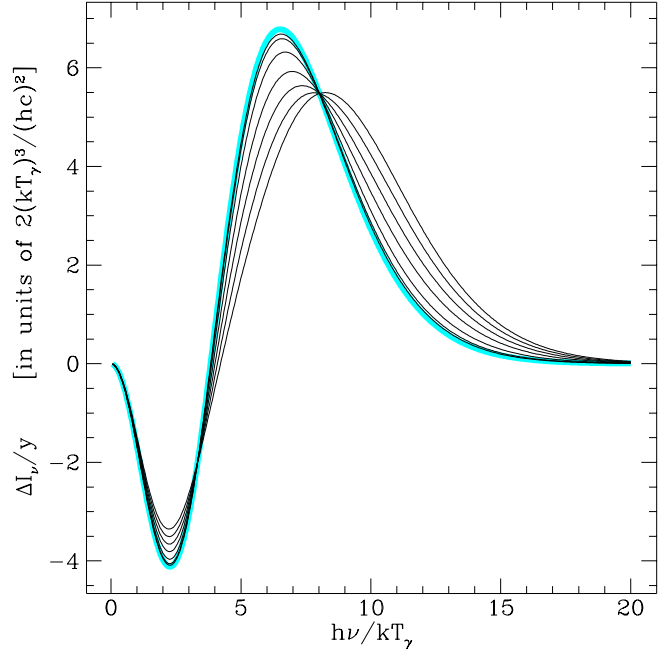}
\caption{Plotted is the small deviation in intensity from a blackbody divided
by the classical $y$-parameter caused when blackbody photons pass through a hot
gas of electrons.  This is computed using the extension of the $y$-distortion
given in the text.  The gray band is centered on the classical $y$-distortion
which applies when $kT_\rme\ll m_\rme c^2$.  The black lines are for electron
temperatures of 1, 2, 5, 10, 15, 20, and 25\ keV.  We have of course assumed
$T_\rme\gg T_\gamma$.  The curves intersect at the zeros of the function
$\Delta n_{\s\rm SZ}^{\s(2,0)}$.}
\label{fig:ExtendedYdistortion}
\end{figure}

	The idea of the Sunyaev Zel'dovich distortion is that one starts out
with a background radiation which is close to a blackbody spectrum, just what
we expect to be produced by the early universe, and it is {\it slightly}
distorted by the action of hot ionized gas through the Compton scattering
process we have just described.  In this small distortion limit we need just
substitute in a blackbody spectrum, $n^{\s\rm BB}$ of eq~\ref{BlackBody}, into
the right-hand-side of the Kompaneets equation.  Let us generalize this idea a
bit by instead considering the more general Fokker-Planck expansion which is an
expansion in $T_\rme$ and $\alpha$.  In this small distortion limit the
different terms will add linearly to the total distortion which we may write as
a sum
\begin{equation}
\Delta n_\gamma=
  \sum_{n\ge0}\sum_{m\ge0}Y_{\s\rm C}^{\s(n,m)}\Delta n_{\s\rm SZ}^{\s(n,m)}(x)
                                    \qquad   x={\epsilon\over k T_\gamma}
\end{equation}
where
\begin{equation}
Y_{\s\rm C}^{\s(n,m)}=\int d\tau\,\left({kT_\rme  \over m_\rme c^2}\right)^n
                                \,\left({kT_\gamma\over m_\rme c^2}\right)^m
\end{equation}
and the superscript ${}^{\s(n,m)}$ correspond to the $\calO(n,m)$ contributions
to the Fokker-Planck expansion.  Substituting $n^{\s\rm BB}(x)$ into the
various terms of eq~\ref{ExtendedKompaneets} we find that
\begin{eqnarray}
&&\hskip-30pt
\Delta n_{\s\rm SZ}^{\s(0,0)}(x)=0                                          \cr
&&\hskip-30pt
\Delta n_{\s\rm SZ}^{\s(1,0)}(x)=\hskip8pt
                {x e^x\over(e^x-1)^2}\,\left(x\,{e^x+1\over e^x-1}-4\right) \cr
&&\hskip-30pt
\Delta n_{\s\rm SZ}^{\s(0,1)}(x)=-{x e^x\over(e^x-1)^2}\,
                                       \left(x\,{e^x+1\over e^x-1}-4\right) \cr
&&\hskip-30pt
\Delta n_{\s\rm SZ}^{\s(2,0)}(x)={x e^x\over(e^x-1)^2}\,
        \Biggl(- 10
              +{47\over 2}x\,{ e^x+1                 \over e^x-1   }
              -{42\over 5}x^2{        e^{2x}+ 4e^x+1 \over(e^x-1)^2}        \cr
&&\hskip108pt +{ 7\over10}x^3{(e^x+1)(e^{2x}+10e^x+1)\over(e^x-1)^3}\Biggr)\ .
\label{ExtendedYdistortion}
\end{eqnarray}
One does expect that to each order in energy that a blackbody spectrum is a
stable solution when the electron and photon temperature are equal so one
should expect the sum rule
\begin{equation}
\sum_{n=0}^N \Delta n_{\s\rm SZ}^{\s(n,N-n)}(x)=0
\end{equation}
and this does seem to be true for $N=0$ and and $N=1$.

	The classical Sunyaev-Zel'dovich $y$-distortion contains only the
$\calO(1,0)$ and $\calO(0,1)$ terms and may be written
\begin{equation}
\Delta n=y\,{x e^x\over(e^x-1)^2}\,\left(x\,{e^x+1\over e^x-1}-4\right)
\label{Ydistortion}
\end{equation}
where
\begin{equation}
y=Y_{\s\rm C}^{\s(1,0)}-Y_{\s\rm C}^{\s(0,1)}
=\int d\tau\,{k(T_\rme-T_\gamma)\over m_\rme c^2}\ .
\label{Yparameter}
\end{equation}
This is the $y$-distortion plotted in fig~\ref{fig:yDistortion} and used in
eqs~\ref{SZdistortion}\&\ref{SZparameter}.  To see how much this classical
formula errs we compare the different expression for a range of electron
temperature in fig~\ref{fig:ExtendedYdistortion}.  We see that the $\calO(2,0)$
corrections become significant when $T_\rme\simgt5$\keV.  This 2nd order
distortion agrees well with the computation of the collision integral by
Rephaeli\cite{Rephaeli95} so higher order corrections do not seem to be
important for $T_\rme\simlt15$\keV.

\section{The Future}

	In the past decade we have witnessed astounding advances in the
measurement of the CMBR spectrum.  After decades of tantalizing evidence of
deviations from a blackbody spectrum we find that the spectrum is amazingly
close to a perfect blackbody.  No longer is it possible to consider a universe
with a very hot inter-galactic medium, or that hydrodynamic forces could have
played a large role in forming the {\it large} scale structure.  There is also
little room for non-equilibrium energetic events in the early universe at
redshifts $<10^7$.  In a way this is most unfortunate.  The thermal equilibrium
state contains the least information - all remnant of events in the universe
before $z\sim10^7$ have been thermalized to nothing, or more precisely to one
number: the temperature.  At the moment we really don't know how to interpret
this number, other than to make a rough comparison to the number of baryons
which is observationally rather ill-determined.  Perhaps some day we will have
cosmogenic theories which will predict the baryon-to-photon ratio with great
accuracy.

	Observationally we are approaching a brick wall which is the Galaxy.
At the present level of sensitivity Galactic contamination from dust and
synchrotron radiation is an important contaminant at all wavelengths.  Galaxy
modeling which makes use of a spectral and spatial structure of the Galaxy
observed at a variety of wavelengths will improve as sensitivities improve
however there will be a limit to how accurately one can subtract off the Galaxy
even given perfect data.  We won't be making observations outside of the
Galactic plane any time soon!

	Yet there is still a lot of room for improvement on the decimeter and
meter scale anisotropies.  Also there is this tantalizing evidence for {\it
negative} spectral deviations ....\footnote{Some people never learn.}.

	Things are not bleak.  In fact spectral distortions of the CMBR is a
rapidly growing field.  Multi-frequency observations is beginning to be the
norm for CMBR anisotropy experiments, and with the CMBR satellites we can
expect literally millions of measurements of the CMBR spectrum in different
directions on the sky.  Admittedly there is a big difference between absolute
measurements of the CMBR flux and differential measurements which are required
for anisotropy since the anisotropy spectral measurements are modulo any DC
spectral distortion.  However it is just his sort of measurement which will
make improved Galaxy subtraction possible.  The spectral information obtained
will tell us mostly about the Galaxy and extra-Galactic radio sources, however
with millions of measurements one can always hope for something a little more
interesting.  Along these lines there is the cluster S-Z effect which is a
rapidly growing field.  With increased sensitivity we can look forward to S-Z
selected cluster catalogs, measurements of radial cluster velocities through
the kinematic S-Z effect, and these studies can work their way down to galaxy
groups and even large scale structure filaments of hot gas. We
can even hope to measure the gas temperature from spectrum if it is hot enough.
In the future we can expect the spectrum and anisotropy measurements to become
increasingly intertwined.

\section{Acknowledgements}

Special thanks to the organizers for an excellent meeting and their infinite
patience.  This work was supported by the DOE and the NASA grant NAG5-2788.

\section{Bibliography}

	What follows is not a list of articles cited in this work, although it
includes all articles cited, but rather an (incomplete) bibliography of
published works related to the CMBR spectrum, including title, listed
alphabetically by the name of the first author.  Many of these papers are of
only historic interest: theories have been ruled out and observations
superseded.  I hope some readers will find it a useful reference.\footnote{The
compilation method was somewhat haphazard and the author apologizes to the
authors of the many important works which are not listed.  Incompleteness is
probably fairly large for papers written in the past decade, and entire subject
areas (e.g. decaying particles) have been omitted. Conference proceedings and
other articles in books have been excluded.}

\end{document}